\documentclass[prc,twocolumn,superscriptaddress,showpacs,preprintnumbers,amsmath,amssymb,f
loatfix]{revtex4}

\usepackage{graphicx}
\usepackage{dcolumn}
\usepackage{bm}
\usepackage{longtable}

\begin{document}

\title{ A Comparison of Shell Model Results for Some properties of the
  Even-Even Ge Isotopes}

\author{S.J.Q. Robinson}

\affiliation{Department of Physics,
Millsaps College,
Jackson, MS, 39210,
}

\author{ L. Zamick}

\affiliation{Department of Physics and Astronomy, 
Rutgers University,
Piscatatway,New Jersey  08854
}

\author{Y.Y. Sharon}

\affiliation{Department of Physics and Astronomy, 
Rutgers University,
Piscatatway,New Jersey  08854
}

\date{\today}

\begin{abstract}

In this work we examine two recent effective shell model interactions,
JUN45 and JJ4B, that have been proposed for use in the
$f_{5/2},p_{3/2}, p_{1/2}, g_{9/2}$ model space for both protons and
neutrons. We calculate a number of quantities that did not enter into the
fits undertaken to fix the parameters of both interactions. In
particular we consider static quadrupole moments (Q's) of excited
states of the even-even $^{70-76}$Ge isotopes, as well as the B(E2)
values in these nuclei. (We have previously studied $^{70}$Zn isotopes
using JJ4B.) Some striking disagreements between the JUN45 prediction
and the experimental results had already been noted for the quadrupole
moments of the $2_1^+$ states of these nuclei. We investigate whether
these discrepancies also occur for the JJ4B
interaction. Subsequently, we also apply both interactions to
calculate the Q's of some more highly excited states and compare the
two sets of predictions regarding the nature of the nuclear states
under consideration. In order to gain insight into these more complex
large-scale shell-model calculations, we examine the corresponding and
much simpler single-j shell model calculations in the $g_{9/2}$
neutron shell.

\end{abstract}

\maketitle

\section{Introduction}

To make shell model calculations tractable one must limit the number
of allowed shell model orbitals that are included. One must then find
a suitable effective interaction in the resulting truncated model
space. One would prefer to construct such an interaction from first
principles. In practice however, one sets the final parameters of a
given interaction by optimizing simultaneously, for many nuclei, fits
to the experimental data for selected nuclear properties (usually the
level excitation energies and the binding energies). This is now
normally done by a process due to Chung and Wildenthal known as
the Linear Combination Method (LC) \cite{chung}. An example using
this procedure can be seen in \cite{gx1}.  When the resulting
interaction is utilized it leads to calculated results for these
selected nuclear properties that are often in very good agreement with
the corresponding measured values.  However, such agreement is not
always obtained for nuclear properties whose data were not utilized in
the fitting of the interaction parameters.

In the present work on the medium mass Germanium isotopes we show that
two such interactions constructed for this region, although very
promising, do not always yield sufficiently accurate results for some
of the nuclear properties.  Indeed, developing a good phenomenological
interaction\ is not a trivial matter. This is especially true for the
T=0 parts of the two-body interactions, parts which are not present for
systems of identical particles. This challenge has been addressed in
\cite{sin1,sin2}.

Previously the current authors showed the importance of including the
$g_{9/2}$ shell in explaining the properties of
$^{70}$Zn\cite{mucher09}.  In that work only one effective
interaction, JJ4B, was used. Here we continue on to the Ge isotopes
using two proposed effective interactions, JJ4B
\cite{mucher09,lis1,lis2,walle1,walle2,recent} and the newer JUN45
\cite{honma09}, which were constructed for the
$p_{3/2}$,$p_{1/2}$,$f_{5/2}$ and $g_{9/2}$ orbitals for both protons
and neutrons. The model space consists of a closed $^{56}$Ni core plus
many valence nucleons.

Our testing ground will be the $^{70,72,74,76}$Ge isotopes, where we
investigate the B(E2)'s and the static quadrupole moment values. These
properties were not considered in fitting the parameters for either
interaction. We also study, to provide contrast, the excitation
energies which were involved in the fitting procedures.

One of the motivations for the present work are the results presented
for the Ge isotopes in Figure 8 of a recent paper by Honma et al
\cite{honma09}. It is seen that for N $\ge$ 38, with the JUN45
interaction, the E($2_1^+$) values are well described, the B(E2; $2_1
\rightarrow 0_1$) values fairly well described, while the Q(2$_1^+$)
values were not in good agreement with the experimental values.  We
were therefore motivated to use the previously-employed JJ4B
interaction of Lisetskiy and Brown ~\cite{lis1} to calculate in the
same space these same nuclear properties. Subsequently we continue
using both interactions to study the excitation energies, B(E2)
values, and Q moments of some more highly excited states in the Ge
isotopes to compare with each other, and whenever possible, with
experimental data.

Finally, to gain insights into the above complex large-scale shell
model calculations, we compute the static quadrapole moments of the
$2_1^+$ states of these Ge isotopes with the simpler single-j shell
model, using only $g_{9/2}$ neutron configurations.

\section{Results}

\subsection{Energies}

Since the excitation energies were used in the fits for the
interaction parameters, we expect that the calculations for them with
the two interactions will yield results in reasonable agreement with
the experimental data. This is indeed the case. For the E($2_1^+$)'s
we see excellent agreement with the JUN45 interaction in the
upper-most right hand part of Figure 8 of Reference \cite{honma09}.
In Table I we present the energies for the J$^{\pi}$= 2$_1^+$,
$0_2^+$, $2_2^+$, and $4_1^+$ states of the Ge isotopes calculated
with both the JJ4B and JUN45 interactions. The fits are pretty good
except for the JJ4B fits for the 0$_2^+$ states.

With JUN45, the average absolute deviation between experimental and
calculated excitation energies is 0.133 MeV. With JJ4B that average
deviation is 0.234 MeV leaving out the $0^+_2$ states and 0.349 MeV if
they are included.  The best fit with JUN45 is for $^{72}$Ge with an
average deviation of 0.068 MeV; with JJ4B the the smallest average
deviation is 0.226 MeV for $^{76}$Ge. If we ignore the possible intruder
$0_2^+$ state, it would be in $^{72}$Ge, at 0.089 MeV.

\subsection{B(E2) Values}

Next we examine the B(E2) values. The calculated values for B(E2)'s
will depend on the effective charges used. We use the standard
$e_p=1.5$ and $e_n=0.5$ values and present our results in Table II for
both the JUN45 and JJ4B interactions. However, these are not always
the preferred values. In \cite{honma09} the values of $e_p= 1.5 e$ and
$e_n=1.1 e$ were used with JUN45 resulting in larger B(E2) values
closer to the experimental values. Values of $e_p= 1.76 e$ and
$e_n=0.97 e$ were used with JJ4B in \cite{mucher09} also putting
calculated values closer to experimental measurements. The
experimental values in Table II are derived from the NNDC database.
In Federman and Zamick \cite{federman69} the calculated neutron
effective charge was larger than 0.5.

Excluding the $2_2^+\rightarrow0_1^+$ transition which has very small
B(E2) values both experimentally and theoretically, the JJ4B values
are almost always bigger by 10 to 30 percent than the JUN45 values.

The experimental value of the B(E2; $2_2^+ \rightarrow 2_1^+$) for
$^{70}$Ge is exceptionally large.  Excluding the very small $2_2^+
\rightarrow 0_1^+$ transition, the experimental B(E2) values are
larger than the JUN45 calculated values for every case and larger than
the JJ4B results in 10 of the 12 cases. These experimental values are
always larger than 17 W.u., indicating some collectivity.

Overall, the calculated values with either interaction are clearly
smaller than the experimentally measured values for $^{74,76}$Ge
(which are very similar experimentally) by an average of about 40
percent indicating an underestimate of the collectivity. This can be
remedied by the use of larger effective charges, a choice which may be
justified because our shell-model space is too small, especially for
the neutrons.

In the calculated B(E2) results with either interaction there is
little change across the Ge isotopes, ranging from about an 8 to 22
percent change for any specific transition. Experimentally more change
is seen.

The experimental value of BE($4_1^+ \rightarrow 2_1^+$)/BE($2_1^+
\rightarrow 0_1^+$) ratios in these nuclei are for $^{70}$Ge 1.14, for
$^{72}$Ge 2.08, for $^{74}$Ge 1.24 and for $^{76}$Ge 1.31.  In a
simple vibrational picture the value of this ratio would be 2.

As is common in LC fit interactions, the B(E2) values were not
included in fitting either interaction's parameters. The experimental
B(E2) values are indeed not fit nearly as well as the energies. With
the standard effective charges, the calculated results with the JJ4B
interaction are closer to the experimental results than the JUN45
interaction. The use of larger effective charges would be a sensible
decision as excitations from the $f_{7/2}$ orbit are not included in
the model space and bring the calculated B(E2) values closer to the
experimental ones.

\subsection{Static Quadrapole Moments}

We now look at the static quadrapole moments of the $2_1^+$, $2_2^+$,
and $4_1^+$ states of the $^{70,72,74,76}Ge$ isotopes.  The measured
and calculated results are presented in Table III. For the quadrupole
moments, experimental results are available only for the $2_1^+$
states. For the $2_2^+$ and $4_1^+$ states we can only compare the
different calculated predictions of the two effective interactions.
Again the proton and neutron effective charges play an important role.
We continue to use in all of our calculations $e_p=1.5$ e and $e_n=0.5$ e.

We begin by comparing the experimental and calculated results for the
Q($2_1^+$)'s. The JUN45 predictions, while in agreement with the
measured results for $^{70}$Ge, are in disagreement for the other
three isotopes. Indeed, for the other three isotopes the experimental
Q(2$_1^+$) values are larger and negative, indicating a prolate
intrinsic shape, while the JUN45 results are positive suggesting an
oblate intrinsic shape.  The JJ4B interaction does a little better
than JUN45. The JJ4B values agree with experiment in the case of
$^{70}Ge$ and $^{76}$Ge but have the opposite sign for $^{72}$Ge and
while of the correct sign are much smaller in magnitude in the case of
$^{74}$Ge. The use of larger effective charges does not resolve the
above discrepancies.

We also calculated quadrapole moments for the 2$_2^+$ state and
4$_1^+$ state even though there are no available experimental
data. Here, we can only compare the results obtained with the two
interactions. For $^{70}$Ge and $^{72}$Ge there is some agreement, the
signs are the same with roughly similar magnitudes. However, the
results are quite different for $^{74}$Ge and $^{76}$Ge. There the
signs are always different between the two interactions except for the
Q($4_1$) of $^{76}$Ge where the signs agree but there is a large
difference in magnitude.

We cannot assess on the basis of Table III which interaction is
better.  The disagreement with experiment for the Q(2$_1^+$) are too
large for both interactions. But our results indicate that more
theoretical work must be done to improve the calculated values of the
quadrapole moments of these excited states of the even Germanium
isotopes. Of course, any experimental measurement of the Q($2_2^+$)
and Q($4_1^+$) would be of great value in clarifying this picture.

It is not clear why there is such a large discrepancy between the
theoretical and experimental Q(2$_1^+$) values, but the results point
out the importance of including data on static quadrapole moments when
fitting the parameters of effective interactions.  We note that in the
simple harmonic vibrational model the static quadruplole moments would
be zero. The results seem to be very sensitive to specific details. We
would guess that the problem is not so much with the two specific
interactions that are used but rather with the specific truncated
shell-model space which is used by both interactions.

From the collective perspective, we note that the ratio of excitation
energies E($4_1^+$)/E($2_1^+$) for the 4 isotopes under consideration
has the respective values of 2.07, 2.07, 2.46 and 2.80 in $^{70}$Ge,
$^{72}$Ge, $^{74}$Ge, and $^{76}$Ge.  In the simple vibrational model
the value of this ratio would be two. Thus the two lighter isotopes
appear to be more vibrational than the two heavier ones. Such a trend
is also present in the experimental Q($2_1^+$) values where the
Q($2_1^+$) of $^{74}$Ge and $^{76}$Ge are larger. The B(E2) ratios
data also appeared vibrational in the case of $^{72}$Ge.

To gain a further perspective for trying to understand the behavior of
the static quadrapole moments of the even Ge isotopes, we consider in
the next section the static quadrapole moments of the Ge isotopes in a
single-j shell model.  More specifically we consider the $g_{9/2}$
neutron subshell. This is not a totally realistic picture. The
wavefunctions that we obtain for the even Ge nuclei in our large-scale
shell model calculations with either JUN45 or JJ4B are very
fractionated, fragmented over many shell model configurations.  This
indicates a more collective, rather than a single particle, picture. For
example with the JUN45 interaction, the "closed shell" configuration
in $^{72}$Ge with J=0 is only 7 percent. However our work does shed
light on what happens to the static quadrapole moments as neutrons
gradually fill a single j shell.

\section{Quadrupole moments in the $g_{9/2}$ shell}

By definition the quadrupole moment is proportional to the expectation
value of the $(2z^2-x^2-y^2)$ operator in the state where the m
projection is equal to j.

In a corresponding semiclassical picture for a single nucleon in a j
shell orbiting outside a closed spherical core, if the orbital angular
momentum vector points along the z axis then the particle is orbiting
in the xy plane. Thus for the ground state of a nucleus with a single
valence nucleon the shape is oblate and the quadrupole moment is
negative for this pancake-like situation.

The formula for the quadrupole moment $Q_{sp}$ of a single nucleon is
a single j shell is

\begin{equation}
Q_{sp} = -\frac{ 2j-1}{2(j+1)} <r^2> \frac{e_{eff}}{e}
\end{equation}

where sp denotes single-particle values, $e_{eff}$ is the effective
charge, and $<r^2>$ is the expectation value of $r^2$ in the
single-particle state.

Aside from the case of j=$\frac{1}{2}$, where the quadrupole moment is
zero, the expression is negative for all half-integer values assuming
a positive nucleon charge e. The expectation value with harmonic
oscillator wave functions is given by

\begin{equation}
<r^2>=(2N+3/2) b^2.
\end{equation} 
Here N is the principle quantum number (N=2n+l where n is the number
of nodes in the radial wave function, not counting r= infinity, and l
is the orbital angular momentum) and $b^2= \hbar/m \omega$, where m is
the nucleon mass and $\omega$ the harmonic oscillator frequency. The
$\hbar \omega$ is usually evaluated as $\hbar \omega= 45/A^{1/3} -
25/A^{2/3}$ or sometimes more simply $\hbar \omega = 41/A^{1/3}$.

The $Q_{sp}$ formula can be generalized \cite{lawson} to the case of n
identical particles in a single j shell, with n odd. It becomes, for
the ground state of an odd nucleus with J=j and seniority 1, $Q= -
\frac{2j+1-2n}{2(j+1)} <r^2> \frac{e_{eff}}{e}$ which for n=1 reduces
to Eqn. (1).

In this simple model, Q is linear in n. As the single j shell fills up
n goes from 1 to (2j+1). As n increases, the quadrupole moment is
negative and of decreasing magnitude till midshell, where Q=0. As n
increases past the midshell, the quadrupole moment becomes increasingly
positive. It thus follows that at midshell Q vanishes and due to the
odd symmetry about the midshell the quadrupole moment of a hole is
minus that of a particle.

Evaluating Eq. 1 for $Q_{sp}$ for a neutron in the $g_{9/2}$ shell
with $e_{n_{eff}}$=1, j=9/2 we find $Q_{sp}= -4 b^2$. With A=72,
$b^2=4.424$ fm$^2$ and $Q_{sp}=-17.696$ $(fm)^2$, a negative value as
expected.

We can next use Racah coefficients to evaluate the values of the
quadrupole moments of the $(g_{9/2})^2$ and $(g_{9/2})^4$ neutron
configurations when these configurations are coupled to a total
angular momentum I of 2 or 4.\cite{yos72} The results, using in the
calculations the same parameters as for $Q_{sp}$, are tabulated in
Table IV both in terms of their values and in terms of $Q_{sp}$. We
see there that the states for $(g_{9/2})^2$ all have seniority v=2
while for $(g_{9/2})^4$ the states can have v=2 or v=4.

For the $(g_{9/2})^4$ configuration there is one special I=4 v=4 state
that is denoted by $v_s=4$ in Table IV. That state is an eigenstate of
any interaction and it does not mix with either the I=4 v=2 state or
the the other I=4 v=4 state \cite{lastz}.

We see from Table IV that Q is positive for both the I=2 and I=4
states of the n=2 case and for both I=2 v=2 and I=4 v=2 states of the n=4
case.  We can associate as a simplistic approximation $^{74}$Ge with
n=2 and $^{76}$Ge with n=4 (all while acknowledging the
fragmented/collective nature of the $^{74}$Ge and $^{76}$Ge calculated
shell-model wavefunctions.) Then the single-j shell model signs for
the Q's disagree with the experimental results but agree with the
signs of the JUN45 results better than with the signs of the JJ4B
results. For the $(g_{9/2})^2$ configuration for I=2, $Q=\frac{-2}{3}
Q_{sp}$ and for I=4, $Q=-0.424 Q_{sp}$.

It is interesting to investigate, for various values of j, the
relationship to Q$_{sp}$ of the quadrupole moment Q(2$_1^+$) of the
$(j)^2$ I=2 v=2 state. These results are given in Table V. Aside from
the case of $j=\frac{3}{2}$ [where Q =0 as the $(3/2)^2$ configuration
  corresponds to the midshell], the ratio of the $Q(2_1)/Q_{sp}$
varies very little, always being negative and ranging from -0.57 to
-0.67.

One might note that the neutron $(g_{9/2})^2$ I=2 v=2 result in Table
IV, $Q(2_1^+) = 11.797 (fm)^2$ is very close to the values in Table
III for $^{72}$Ge in the large shell model calculations. This is true
regardless of which of the two interactions one considers.  This
cannot however be taken seriously. The calculated wavefunctions are
highly fractionated and, furthermore, in the $g_{9/2}^2$ neutron
configuration the magnetic moment would be negative (-0.425) but from
\cite{stone05} it is known that in $^{72}$Ge the measured magnetic
moment is positive.  One possible explanation is that a p$_{1/2}$
particle has no quadrupole moment.  So if the low-lying configurations
are dominated by p$_{1/2}$ and g$_{9/2}$ neutrons, then only the
g$_{9/2}$ would contribute.

In the simplest shell model picture $^{72}$Ge consists of a closed
proton shell and a closed neutron shell. The last occupied orbits are for
the protons p$_{3/2}$ and for the neutrons p$_{1/2}$. One simple
configuration for the $2^+$ state would be to promote 2 neutrons from
p$_{1/2}$ to g$_{9/2}$. In this approximation the results of Table
IV would apply.

\section{Summary}

In this paper, using the even Ge isotopes, we have called attention to
the importance of trying to include as many nuclear properties as
possible when fitting the residual effective interaction
parameters. The excitation energies, which were included in such fits,
can be calculated well. On the other hand, the B(E2) values and the
$Q(2_1^+)$ values were not included in the fits for the interaction
parameters. Their calculated values, especially for the quadrupole
moments, are shown to differ substantially from their measured values.

The single-j shell model provides some physical insights into how the
static quadrupole moments behave in a simple model as the j shell
occupation increases.

The authors would like to thank Dr. N. Benczer-Koller for her interest
in this work.

\begin{table}
\begin{center}
\caption{Excitation Energies in MeV for the even-even Germanium
  isotopes. Experimental values taken from the NNDC database.}
\label{tab:table1}
\begin{tabular}{| c | c | c | c | c |} \hline
     & $^{70}$Ge &$^{72}$Ge&$^{74}$Ge& $^{76}$Ge \\ \hline
E($2_1^+$) &  & &  &  \\ \hline
Experiment & 1.039 &0.834 & 0.596 & 0.563 \\ \hline
JJ4B & 0.737 &0.710 & 0.737  & 0.718 \\ \hline
JUN45 & 0.907 & 0.814 & 0.717  & 0.745 \\ \hline
       & & & &  \\ \hline
E($0_2^+$) &  & &  &  \\ \hline
Experiment & 1.216 &0.691 & 1.483 & 1.911 \\ \hline
JJ4B & 1.952 &2.025 &1.937  & 2.162 \\ \hline
JUN45 & 1.084 &0.761 &1.461  & 1.995 \\ \hline
       & & & &  \\ \hline
E($2_2^+$) &  & &  &  \\ \hline
Experiment & 1.708 &1.464 & 1.204 & 1.108 \\ \hline
JJ4B & 1.347 &1.351 &1.371  & 1.368 \\ \hline
JUN45 & 1.404 &1.375 &1.351  & 1.364 \\ \hline
       & & & &  \\ \hline

E($4_1^+$) &  & &  &  \\ \hline
Experiment & 2.153 &1.728 & 1.464 & 1.410 \\ \hline
JJ4B & 1.870 &1.698 &1.735  & 1.653 \\ \hline
JUN45 & 2.027 & 1.820 & 1.613  & 1.637 \\ \hline
       & & & &  \\ \hline
\end{tabular}\end{center}
\end{table}

\begin{table}
\begin{center}
\caption{B(E2) reduced transition strength in W.u. Effective charges
  $e_p=1.5$ $e_n=0.5$ were used. Experimental values were taken from
  the NNDC database.}
\label{tab:table2}
\begin{tabular}{| c | c | c | c | c |} \hline
 & $^{70}$Ge &$^{72}$Ge&$^{74}$Ge& $^{76}$Ge \\ \hline
BE($2_1 \rightarrow 0_1$) & & & &   \\ \hline
Experiment & 20.9(4) & 17.8(3) & 33.0(4) & 29(1) \\ \hline
JJ4B & 19.68 & 19.88 & 19.90  & 18.24 \\ \hline
JUN45 & 14.47 & 14.55 & 16.59  & 16.36 \\ \hline
& & & &  \\ \hline
BE($2_2 \rightarrow 2_1$) & & & &\\ \hline
Experiment & 114(5) &62(+9 -11) & 43(6) & 42(9) \\ \hline
JJ4B & 26.55 &29.34 & 29.17  & 22.94\\ \hline
JUN45 & 23.48 & 24.70 & 24.88  & 25.38 \\ \hline
& & & &  \\ \hline
BE($4_1 \rightarrow 2_1$)  & & & &\\ \hline
Experiment & 24(7) & 37(5) & 41(3) & 38(9) \\ \hline
JJ4B & 28.22 & 27.62 & 27.04  & 24.15 \\ \hline
JUN45 & 23.65 & 25.08 & 23.46  & 22.04 \\ \hline
& & & &  \\ \hline
BE($2_2 \rightarrow 0_1$) & & & & \\ \hline
Experiment & 0.9(+4-8) &0.130 (+18 -24) & 0.71 (11) & 0.90(22) \\ \hline
JJ4B & 1.67 & 1.37 & 0.12  & 0.01 \\ \hline
JUN45 & 0.71 & 1.21 & 1.35  & 0.42 \\ \hline
& & & &  \\ \hline
\end{tabular}\end{center}
\end{table}

\begin{table}
\begin{center}
\caption{Static quadrupole moments in (fm)$^2$. Effective charges of
  $e_p=1.5$ and $e_n=0.5$ were used. N/A indicates unavailable
  data. The experimental data is from \cite{stone05}.}
\label{tab:table3}
\begin{tabular}{| c | c | c | c | c |} \hline
 & $^{70}$Ge &$^{72}$Ge&$^{74}$Ge& $^{76}$Ge \\ \hline
Q($2_1^+$) &  & &  &  \\ \hline
Experiment & 3(6) or 9(6) & -13(6) & -25(6) & -19(6) \\ \hline
JJ4B & 15.13 &10.97 & -5.89  & -14.50 \\ \hline
JUN45 & 9.94 & 12.85 & 12.02  & 1.77 \\ \hline
       & & & &  \\ \hline
Q($2_2^+$) & & &  &  \\ \hline
Experiment &N/A  &N/A &N/A  &N/A  \\ \hline
JJ4B & -15.42 & -11.31 & 5.37  & 15.49 \\ \hline
JUN45 & -13.27 &-13.48 & -11.53  & -0.06 \\ \hline
       & & & &  \\ \hline
Q($4_1^+$) &  & &  &  \\ \hline
Experiment &N/A  &N/A &N/A  &N/A  \\ \hline
JJ4B & 3.16 &3.15 &-8.30  & -14.03 \\ \hline
JUN45 & 1.36 &8.50 &11.29  & -1.32 \\ \hline
       & & & &  \\ \hline
\end{tabular}\end{center}
\end{table}

\begin{table}
\begin{center}
\caption{Calculated static quadrupole moments in the single-j shell
  model space for the $g_{9/2}$ neutrons. Here n is the number of particles,
  I the total angular momentum, and v the seniority.  For n=1
  $Q_{sp}=-4b^2=-17.7 (fm)^2$ (see text)}
\label{tab:table4}
\begin{tabular}{| c | c | c | c |} \hline
n=1 & $Q_{sp} = -4 b^2 = -17.7 fm^2$ &  &   \\ \hline
  &   & $\frac{Q}{Q_{sp}}$ & $\frac{Q}{e} (fm)^2$ \\ \hline
n=2 & I=2 v=2 & -2/3 & 11.797 \\ \hline
    & I=4 v=2 & -0.424 & 7.686 \\ \hline
n=4 & I=2 v=2 & -0.221 & 3.913 \\ \hline
    & I=2 v=4 & 0.129& -2.279 \\ \hline
    & I=4 v=2 & -0.141 & 2.502 \\ \hline
    & I=4 $v_s$=4& -0.751 & 13.287 \\ \hline
    & I=4 v=4 & 0.495  & -8.767 \\ \hline
\end{tabular}\end{center}
\end{table}

\begin{table}
\begin{center}
\caption{For various $(j^2)^{I=2, v=2}$ configurations, the
  relationship of $Q/Q_{sp}$ for that j value.}
\label{tab:table5}
\begin{tabular}{| c | c | c | c |} \hline
n=2 &   & j & $\frac{Q}{Q_{sp}}$\\ \hline
    & I=2 &      &  \\ \hline
    &     & 3/2   & 0 \\ \hline
    &     & 5/2   & -0.5714 \\ \hline
    &     & 7/2   & -0.6531 \\ \hline
    &     & 9/2   & -0.6667 \\ \hline
    &     & 11/2   & -0.6649 \\ \hline
    &     & 13/2   & -0.6593 \\ \hline
    &     & 15/2   & -0.6530 \\ \hline
\end{tabular}\end{center}
\end{table}

  \bibliography{Sources}

\begin{thebibliography}{10}

\bibitem{chung} W. Chung, Ph.D. thesis, Michigan State University, 1976.

\bibitem{gx1} M. Honma et al., Nucl. Phys. A704, 134c (2002).

\bibitem{sin1} J. Sinatkas, L.D. Skouras, D. Strottman and
  J.D. Vergados, J. Phys. G. Nucl. Part. Phys. 18, 1377 (1992).

\bibitem{sin2} J. Sinatkas, L.D. Skouras, D. Strottman and
  J.D. Vergados, J. Phys. G. Nucl. Part. Phys. 18, 1401 (1992).

\bibitem{mucher09} D. Mucher et. al., Phys. Rev. C 79, 054310 (2009).

\bibitem{lis1} A. F. Lisetskiy and B.A. Brown, private communication.
  (2010).

\bibitem{lis2} A.F. Lisetskiy, B.A Brown, M. Horoi and H. Grawe,
  Phys. Rev. C 70, 044314 (2004).

\bibitem{walle1} J.Van de Walle et al., Phys. Rev. Lett. 99, 142501
  (2007).

\bibitem{walle2} J. Van de Walle et al., Phys. Rev. C 79, 014309 (2009).

\bibitem{recent} B. Cheal et. al., Phys. Rev. Lett. 104, 252502
  (2010).

\bibitem{honma09} M. Honma, T. Otsuka, T. Mizusaki, and
  M. Hjorth-Jensen, Phys. Rev. C 80, 064323 (2009).


\bibitem{lawson} R.D. Lawson, Theory of the Nuclear Shell Model,
  Clarendon Press, Oxford (1980).

\bibitem{stone05} N.J. Stone, At. Data Nucl. Data Tables 90, 75
  (2005).

\bibitem{federman69} P.Federman and L.Zamick, Phys. Rev. 177, 1534
  (1969).

\bibitem{yos72} S. Yoshida and L. Zamick, Ann. Rev. of Nucl. Sci. 22,
  121 (Palo Alto) (1972).

\bibitem{lastz} L. Zamick and P. van Isacker, Phys. Rev. C 78, 044327
(2008) and references therein.


\end{thebibliography}

\end{document}